\documentclass[10pt]{article}

\usepackage{cite} 

\usepackage{graphicx}
\usepackage[font=scriptsize]{caption}

\usepackage[utf8]{inputenc} 

\begin{document}

\title{\textbf{Researcher for one day: an Astrophysics Masterclass}}
\date{}
\maketitle

\begin{center}
\author{I De Angelis$^{1,2}$, A Postiglione$^{1,2}$, F La Franca$^{1}$  }\\
\end{center}

\paragraph{} \parbox[t]{1\columnwidth}{$^1$Dipartimento di Matematica e Fisica, Universit\`a degli Studi Roma \\Tre, Rome (ITALY)\\%
    $^2$INFN Sezione di Roma Tre, Rome, (ITALY)\\
    
    ilaria.deangelis@uniroma3.it}

\begin{abstract}
In the last years it is becoming more and more evident the importance in providing to the students, and people in general, a more realistic view on the scientific research world. In this framework many initiatives are beeing carried out to put students closer to scientists.
The Astrophysics Masterclass, held in our  Department, is an example of such activities, consisting in a one-day outreach event during which high school students experience the research methods in the astrophysics field. In this paper we describe this activity, which allows the participants to work on real research data in a real research center.  
\end{abstract}


\section{A day as a researcher}
In recent years, many initatives have been carried out by Universities and Research Institutions in order to reduce the discrepancy between what science really is and the way it is perceived during, and after, school. One example consists in the International Masterclasses \cite{ref:Particle_Physics_Masterclasses, ref:New_Directions}, one-day outreach events during which high school students discover particle physics. At the Department of Mathematics and Physics of Roma Tre University, we developed an analogue activity which uses the key elements of these Masterclasses but based on the astrophysics research carried out in our Department.  

Our activity takes place in the Computer Lab of our Department and involves a group of about 40 high school students, 16-18 years old. Typically the participants have  already dealt with basic physics topics (Newton’s laws of motion, concepts of speed, energy and power) and basic math (first degree equations, logarithms). They are selected by their own teachers, by taking into account their expression of interest. During the astrophysics Masterclass, the students spend one whole day in the  Department, becoming scientists for one day by carrying out research studies on galaxies and black holes. Indeed, once the
scientific objective is presented, the participants are allowed to look at the images captured by telescopes and to analyse real data in order to eventually obtain the results.
The final objective of their research consists in characterizing some galaxies by measuring their brightness, distance, luminosity and mass, until recognizing among them a particular type of galaxy: a galaxy whose luminosity is orders of magnitude larger than normal, thanks to an active accreting supermassive black hole in its center. 
In order to achieve this goal, data and results belonging to a real research project, partly carried out in the framework of a PhD thesis \cite{ref:fabio}, are used. The data were obtained using some among the most important (largest) existing astrophysical, ground based and space, international observing facilities. In this study, the sky region European Large Area ISO Survey S1 (ELAIS-S1) is analyzed thanks to observations made in the visible wavelengths by the Very Large Telescope (VLT/ESO), in the infrared wavelengths by the NASA satellite Spitzer and in X-ray wavelengths by the ESA satellite XMM-Newton.

Within this sky region, four astrophysical sources have been chosen: a galaxy in which there is a strong star formation in place (a \textit{starbust}); a galaxy in which there is no relevant star formation (a \textit{passive galaxy}); and two \textit{Active Galactive Nuclei} (AGNs), i.e. galaxies with a much-higher-than-normal luminosity nucleus powered by accreting matter onto a Super Massive Black Hole. As for one AGN it is not easy to study its centre, it is called obscured AGN or type 2 AGN \cite{ref:Antonucci}; the other AGN is instead of type 1, i.e its spectrum shows the broad emission lines produced near the black hole, and therefore the black hole mass and its accretion rate can be measured. Each source was characterized through measuring their flux, distance, redshift, luminosity, and number of stars.
In order to allow the students to work directly on these sources, simplified procedures have been built, based on professional astrophysical data analysis softwares, such as MIDAS of the European Southern Observatory or, in recent years, Python codes.  

\section{The steps of our Masterclass}
The Masterclass in Astrophysics lasts one day and is characterized by: an education phase, a training phase, a test phase, and eventually a discussion and comment of the results. In designing the activity we paid particular attention to let the participants to feel fully immersed in the scientific research world. The students are also encouraged
to take advantage of some time-breaks, during which impressions, information, curiosities and doubts can be shared with more experienced people, like PhD or undergraduate students. 
Moreover, with this last objective in mind, an observation, in the morning, of the Sun at the outside telescope, and a common lunch at the University canteen, together with our tutor students, are organized.
During the activity, particolar attention is dedicated to the contextualization of the research beeing carried out, and to its links to the big open questions of physics, to the role of the current instrumentation and to the perspectives of the research field. In our opinion, this strengthens the idea of science as a continuous evolving process of discoveries and knowledge, and not as mere set of sophisticated laws.

\subsection{Education Phase}
Our activity starts in the morning with a presentation on the theoretical foundations of the research the participants will carry on. 
The most important astrophysical open questions are illustrated, in order to allow the participants to understand the role of their research in the long path of knowledge. 
Two lectures, held by professors or senior researchers, introduce what we know about our Universe, its expansion and the objects that populate it (galaxies, Active Galactic Nuclei, black holes). The physical quantities that will be measured during the day are then described: distances and redshifts, luminosities and fluxes, masses, spectra, absorption and emission spectral lines. 
Then the instrumental aspects are discussed. The most important existing telescopes are presented, showing their characteristics and their planned upgrades, focusing on those used during the Masterclass (VLT/ESO, Spitzer, XMM-Newton). During the morning break, students are invited to join the observation of the Sun, so that they can experience the use of a telescope.
Then the lunch is had all together in the University canteen, thus offering the opportunity to discuss with other participants and with University students, both on the activity of the day and on the University life.
The afternoon starts introducing the participants to the scientific problem they have to deal with, and to the tools and methodologies that will be used:  by analyzing some images of a sky region taken by three telescopes, in three different wavelenght bands (mid-infrared, optical and X-rays), the aim is to characterize four astrophysical sources within.

\subsection{Training Phase}
During the training phase participants are divided into couples, each in front a computer, with the task of fully characterizing one of the four astrophysical sources in the celestial field. The work is organized as follows: a senior researcher guides all participants step by step, showing the procedures, while each of the tutors (young PhD students or University students) follow more closely the work of four groups, in order to solve any technical problems or answer to questions. Before moving to each of the next steps, first the tutors verify that all the groups correctly carried out all the required operations and then tell to the senior researcher that it is possible to continue.

The procedure starts with the measure of the flux of the source. Actually the flux was previously measured by the researchers of the Department, while the participants can discover it just by clicking on the image of the source; then, the visible spectrum obtained at the VLT/ESO is shown and analyzed. Comparing its absorption or emission lines with that of similar but local and well known galaxies, it is possible to derive its redshift.  Once the redshift of the galaxy is obtained, it is then possible to calculate its distance. Thus, by combining the information of the flux with that of the distance, the intrinsic luminosity of the object can be obtained, leading to an estimate of the number of the stars populating the galaxy (which results to be about 100 of billions).

\subsection{Obtaining the results}
Once the first source has been analyzed under the guidance of the researcher, the students independently, but with the help of the tutors, characterize the other three sources of the celestial field. It is in this moment that the students use everything they have learned during the day, and catch a glimpse of what scientific research is: dealing with technical problems, conceptual difficulties, mathematical obstacles, but also experiencing the euphoria of the discovery. Eventually all the participants come to the identification of the type 1 AGN, where it is possible to observe the accreting black hole emission. 
At this point, the researcher returns to guide all the participants. A discussion begins on what are the features of the AGN; the spectrum is analyzed, highlighting the width of its emission lines. Eventually, by combining the luminosity value with  the width of the lines, the mass of the central black hole (about 100 million times that of the Sun) and its accretion rate (few solar masses per year)  are estimated. The scientific research objective declared at the beginning of the day is thus completely reached.
The day ends with a debate on the importance of communicating the results to the scientific community. In order to strengthen this idea, a document is distributed to the students, representing the scientific paper that describes the research activity of the day. All the participants are listed among the authors, since they actually had become scientists for one day.

\section{Participants and reactions}
Our astrophysics Masterclasses started in April 2008 with 32 students and then consolidated in the International Year of Astronomy (2009) during which we managed to organize three editions. Since then they have been offered every year.  The participants are usually enthusiastic, as also confirmed by the huge number of applications. Some of them
come from students to whom the activity was recommended by previous participants. 
Recently the number of applications has increased so much that we have decided to offer more than one edition in the same year. To date, a total of 568 students have participated to our Masterclasses
in Astrophysics.

\section{Summary and conclusion}
In this paper we presented an educational activity dedicated to high school students that follows the typical structure of the Masterclass in Particle Physics but instead focuses on Astrophysics. 
We believe this project can efficiently contribute to communicate to young people how science works, and therefore,  by serving as an orientation activity, encourage STEM careers.
Indeed, some of the students who  are attending (have attended) our Degree Course in Physics have previously attended our astrophysics Masterclass. 
Compared to other outreach activities dedicated to schools, these Masterclasses offer the opportunity to better experience scientific research since it allows students to directly work on real scientific data, even if in a simplified manner. Beeing carried out within a University, this Masterclass allow the student to see where scientific research really takes place, bringing them closer to the figure of the researcher.
Moreover, through the contextualization of the research, the presentation of its open questions, it reinforces the idea that science is a process rather than a result, that it is not a set of well defined laws, but instead a discipline in a constant developing status, where each discovery opens new questions and creates new challenges that continually force us to improve our skills. 

The astrophysics Masterclasses are now included in the list of activities for schools which our Department offers every year.
We plan in the future to quantify the feedbacks received from the participants by testing the effectiveness of our activity through the distribution of questionnaires to get information on their satisfaction and the influence that the Masterclass had on their ideas about science and the scientific careers.

\section*{Acknowledgments}
This work was supported by the Italian Project ``Piano Lauree Scientifiche". Special thanks go to IAU and UNESCO for having organized the International Year of Astronomy in 2009. The authors also thank the high school teachers who promoted our activity among their students.


\begin{thebibliography}{0}
\bibitem{ref:Particle_Physics_Masterclasses}
Cecire K., 2011, ``Particle Physics Masterclasses", \textit{Proceedings of the DPF-2011 Conference, August 8–13, 2011}.

\bibitem{ref:New_Directions} 
Cecire K. and Dower R., 2019, ``New Directions in International Masterclasses",  \textit{arXiv:1910.0052}

\bibitem{ref:fabio}
Sacchi N., La Franca F., Feruglio C. et al., 2009, ``Spectroscopic Identifications of Spitzer Sources in the SWIRE/XMM-Newton/ELAIS-S1 Field: A Large Fraction of Active Galactic Nucleus with High $F(24 \mu m)/F(R)$ Ratio", \textit{ApJ}, \textbf{703} 1778.

\bibitem{ref:Antonucci}
Antonucci R., 1993,  ``Unified Models for Active Galactic Nuclei and Quasars",  \textit{ARA\&A}, \textbf{31} 473

\end{thebibliography}
\end{document}